# Cooling Dynamics in Multi-fragmentation processes

T.X. Liu, W.G. Lynch, M.J. van Goethem, X.D. Liu, R. Shomin, W.P. Tan, M.B. Tsang, G. Verde, A. Wagner[1], H.F. Xi[2], H.S. Xu[3],

*National Superconducting Cyclotron Laboratory and Department of Physics and Astronomy, Michigan State University, East Lansing, MI 48824, USA,*

W. A Friedman,

*Department of Physics, University of Wisconsin, Madison, WI 53706, USA,*

S.R. Souza, R. Donangelo,

*Instituto de Fisica, Universidade Federal do Rio de Janeiro*

*Cidade Universitária, CP 68528, 21945-970 Rio de Janeiro, Brazil*

L. Beaulieu, B. Davin, Y. Larochelle, T. Lefort, R. T. de Souza, R. Yanez, V. E. Viola,

*Department of Chemistry and IUCF, Indiana University, Bloomington, IN 47405, USA,*

R.J. Charity, and L.G. Sobotka,

*Department of Chemistry, Washington University, St. Louis, MO 63130, USA*

**Abstract**

Fragment energy spectra of neutron deficient isotopes are significantly more energetic than those of neutron rich isotopes of the same element. This trend is well beyond what can be expected for the bulk multi-fragmentation of an equilibrated system. It can be explained, however, if some of these fragments are emitted earlier through the surface of the system while it is expanding and cooling.

---

[1] Institut für Kern- und Hadronenphysik, Forschungszentrum Rossendorf, D-01314 Dresden, Germany.
[2] Present address: Benton Associates, Toronto, Ontario, Canada.
[3] Institute of Modern Physics, Lanzhou, China

The yields of particles emitted in intermediate energy [1,2], relativistic [3,4] and ultra-relativistic [5] nuclear collisions have been successfully compared to equilibrium statistical ensembles. This suggests the attainment of chemical equilibrium in such collisions, motivating investigations [6-8] of phase transitions in strongly interacting matter in systems where stationary thermodynamic equilibrium may not be achieved. Such interpretations assume that the relevant degrees of freedom have time to equilibrate, and that the observables being described reflect that equilibrium.

As in the case of the early universe, observables probing such degrees of freedom in nuclear collisions may reflect a freezeout time when they cease to evolve. The yields of particles emitted per central heavy ion collision have been described successfully, for example, by equilibrium calculations for an expanding nuclear system at a freezeout density of $\rho \approx 1/6\rho_0 - 1/3\rho_0$ [1,4,7]. Careful examinations of freezeout observables may reveal non-thermal details of the freezeout configuration. In the case of nuclear systems, such examinations can identify particles that escape through the surface of the system before it undergoes bulk disassembly and most of its chemical observables freeze out. In this paper, we show how the energy spectra of isotopically resolved fragments can allow one to quantify such effects and assess the accuracy of a global freezeout approximation.

The expected properties for the energy spectra of fragments from the disassembly of thermalized freezeout configuration are straightforward. Because nuclear interactions between particles are assumed to be negligible after freezeout, the observed mean kinetic energies of particles $\langle E_k \rangle$ reflect the thermal kinetic energy $3/2T$, the collective velocity $v_{coll}$ at freezeout and the energy $<E_{Coul}>$ gained due to the accelerations of these particles by the Coulomb field of the remaining system [9-11]:

$$\langle E_k \rangle \approx \frac{3}{2}T + \frac{1}{2}Am_N \langle v_{coll}^2 \rangle + \langle E_{Coul} \rangle, \qquad (1)$$

where $Am_N$ is the mass of the fragment with $A$ nucleons and $m_N$ is the nucleon mass. As the Coulomb energy depends nearly linearly on the fragment charge and for light fragments, $Z \approx A/2$, Eq. 1 suggests a roughly monotonic dependence of $\langle E_k \rangle$ upon $A$; this has been used previously to extract values for the collective expansion velocity $v_{coll}$ after constraining $<E_{Coul}>$ with assumptions about the breakup density [9-11].

Because $E_{Coul}$ depends on Z while the collective motion term depends on A, one might minimize sensitivity to Coulomb effects and isolate the collective motion term by comparing the mean energies of isotopes of each element. However, such studies, performed previously for light charge particles Z≤2, observed mean energies for $^3$He that are higher than those for $^4$He, contrary to Eq. 1 [12,13]. These observations could be reproduced by assuming some emission of $^3$He particles through the surface of the emitting source prior to a thermalized bulk disintegration and freezeout [12,13]. Since early emission of light charged particles via dynamical mechanisms can be expected on general grounds, such observations provide little guidance to the emission mechanisms of heavier fragments. To probe such issues, one should measure isotopically resolved energy spectra for fragments heavier than He.

In our experiment, $^{112}$Sn beams produced from the K1200 Cyclotron at the National Superconducting Cyclotron Laboratory at Michigan State University bombarded a $^{112}$Sn target of 5 mg/cm$^2$ areal density. Nine telescopes of the Large Area Silicon Strip Array (LASSA) [14] detected isotopically resolved particles with 1≤Z≤10. Each telescope consists of one 65 μm single-sided silicon strip detector, one 500 μm double-sided silicon strip detector and four 6-cm thick CsI(Tl) scintillators. The calibration of these telescopes to an accuracy of better than 3% was achieved using alpha sources, elastic scattering and direct fragmentation beams [15].

The center of the LASSA device was located at a polar angle of θ=32° with respect to the beam axis, covering laboratory polar angles of 7°≤ θ ≤ 58° with an angular resolution of about ±0.43°. The multiplicity of charged particles [14] measured with LASSA and the 188 plastic scintillator - CsI(Tl) phoswich detectors of the Miniball/Miniwall array [16] provided impact parameter selection. In the analyses described below, central collisions, corresponding to a reduced impact parameter of $b/b_{max}$ ≤ 0.2 [15] ($b_{max}$ ≈7.3 fm) were selected by a gate on the top 4% of the charged-particle multiplicity distribution. Such central events display many attributes consistent with bulk multifragmentation following expansion and spinodal decomposition at densities of $\rho \leq 1/3\rho_0$ [1,9,18].

Center of mass energy spectra for $2 \leq Z \leq 8$ were obtained by averaging over center of mass angles of $70° \leq \theta_{CM} \leq 110°$. At these angles, the coverage of the LASSA array is excellent, with losses only for fragments emitted at very low energies E/A< 0.2 MeV in the center of mass, corresponding to small laboratory angles of $\theta_{lab} \leq 7°$. The data presented below have been corrected for these efficiency losses and for multiple hits in the detector telescopes.

In Fig. 1, we show the energy spectra for $^{11}$C (open circles) and $^{12}$C (closed circles). The yield of $^{12}$C yield is nearly a factor of 10 higher than that for $^{11}$C reflecting its higher binding energy. The peak in the energy spectrum occurs at higher energies for $^{11}$C than for $^{12}$C and is broader. These two factors dictate a higher mean energy for $^{11}$C than for $^{12}$C.

In the left panel of Fig. 2, the measured mean energies are plotted as a function of the mass number A using the same symbol for isotopes of the same element. The even Z (Z=2, 4, 6, 8) elements are represented by closed symbols and the odd Z (Z=1, 3, 5, 7) elements by the open symbols. Generally, $\langle E_{CM} \rangle$ increases with A; however, the lightest isotopes in each element ($^3$He, $^6$Li, $^7$Be, $^{10}$B, $^{11}$C, $^{13}$N) display a significantly larger mean energies than the neutron rich isotopes. (The measured mean energy of the neutron deficient oxygen isotope, $^{15}$O, is very uncertain due to insufficient statistics and is not plotted.) This striking trend is contrary to that expected for fragment emission from a single equilibrated source and described by Eq. 1.

Because sequential decay reduces the mass of the excited fragments produced at breakup, the trend for an equilibrated system can differ from Eq. 1. To quantify such effects, we modeled equilibrium breakup configurations composed of excited fragments and light particles with the improved Statistical Multi-fragmentation Model (ISMM) of ref. [19] and calculated the subsequent decay of the excited fragments with a Monte-Carlo Weisskopf model [20]. In these calculations, we assumed an initial source with $A_0$=168, $Z_0$=75 and $E^*/A$ = 6 MeV (roughly 75% of the total mass, charge and energy of the combined projectile and target [21]). We assigned randomly an initial thermal velocity to each fragment and light particle according to a Boltzmann distribution characterized by a breakup temperature of $T= 5.24$ MeV [19] and positioned each particle or fragment randomly within a volume of 8.3 fm. We added a collective velocity $\vec{v}_{coll} = \frac{\vec{r}}{R} v_{coll,max}$ to the thermal velocity, solved the classical equations of motion these particles interacting via Coulomb forces, and varied $v_{coll,max} = \sqrt{2 \cdot (E_{coll}/A)/m_N}$ (i.e. $E_{coll}/A$) to describe the mass dependence of the experimental mean energies.

The optimal choice of collective velocity depends on details of the placement of fragments within the volume. If one excludes initial configurations with any part of any fragment outside R, the accepted configurations fragments will be more concentrated in the interior than if one excludes only the configurations with the center of mass of any fragment outside R. The fragments in the former case experience larger Coulomb forces on average than in the latter case; thus, the collective energy per nucleon needed to reproduce the <$E_k$> data, $E_{coll}/A \approx 0$, is less than that for the latter case, $E_{coll}/A \approx 2$ MeV.

As a typical example, we show the predicted $^{11}$C and $^{12}$C spectra (normalized to the data) for the latter calculation in Fig. 1 as the solid histograms. The calculation reproduces the peak of the $^{12}$C spectrum better than the peak of the $^{11}$C spectrum; neither the $^{11}$C nor the $^{12}$C calculation reproduces the high-energy tails of the corresponding experimental spectrum. The right panel in Fig. 2 shows the corresponding predicted mean energies. Slightly reduced values for $\langle E_k \rangle$ are calculated for symmetric N=Z fragments; these reductions reflect strong secondary decay contributions to the yields of N=Z nuclei from the secondary decay of heavier particle unbound nuclei that have somewhat smaller initial velocities. The strength of these secondary decay contributions reflect the Q-value for secondary decay, which favors decays to well bound N=Z nuclei and suppresses decays to their N<Z neighbors. Thus, the calculation predicts slightly larger values of $\langle E_k \rangle$ for N<Z nuclei than for N=Z nuclei, but the calculated change in $\langle E_k \rangle$ is much smaller than that observed experimentally. We therefore conclude that the enhancement in the measured $\langle E_k \rangle$ for the N<Z fragments cannot be attributed to secondary decay. Changing the assumptions about the spatial distribution of fragments or the collective velocity at breakup does not change this conclusion.

This failure and the aforementioned evidence for surface emission of helium isotopes suggest that fragments may also be emitted through the surfaces of the system while it is expanding and cooling. Because the binding energies of the neutron deficient isotopes are significantly less than those of their neutron rich neighbors, their surface emission rates will decrease relative to their neutron rich neighbors as the system cools. The Expanding Emitting Source (EES) model [22] provides means to test this scenario because it provides a self-consistent calculation of emission rates for each isotope from a thermalized system while it is cooling and expanding.

To illustrate these ideas, we have performed an EES calculation, which assumes that particles can be radiated from the surface of the expanding system prior to bulk disintegration and during bulk disintegration itself. Unlike equilibrium models, which assume the system to have already expanded, this time dependent model calculates the expansion and emission of the system beginning at an earlier time as it expands through saturation density. For $\rho \geq 0.4\rho_0$, the model specifically assumes surface emission bulk emission for $\rho \leq 0.3\rho_0$ and a gradual transition from surface to the bulk emission for densities, $0.4\rho_0 > \rho > 0.3\rho_0$. For our EES calculations, we take saturation density, $E^*/A$=9.5 MeV, $A_O$=224 and $Z_O$=100 as the specific starting conditions.

As one example of the EES model results, we show the time evolution of $^{11}$C and $^{12}$C yields. At the time of emission, the primary fragment of each isotope acquires an average kinetic energy dictated by the instantaneous Coulomb barrier, expansion velocity and temperature of the expanding system. (The early surface emission contributions increase the value of the Coulomb and collective contributions above those obtained from the bulk emission alone.) Taking this time dependence into account, we plot the $\langle E_k \rangle$ values for $^{11}$C (dashed line) and $^{12}$C (solid line) as a function of the time of emission in the left panel of Fig.3. Over the evolution of the source the carbon isotopes are emitted with a range of kinetic energies but there is very little difference between the values for the two carbon isotopes at any given time.

We next examine the emission rates as a function of time for the two isotopes. This is shown in the lower right panel of Fig. 3 where the instantaneous rates for each isotope are plotted and in the upper right panel of Fig. 3 where the ratio of the rates is plotted. Here we find that the rate for the emission of $^{11}$C relative to that of $^{12}$C changes drastically with time. The emission of the former is enhanced at earlier times and the latter at later times. Within the EES model, the rates are determined primarily by three factors: the spin degeneracy factors of each channel, free energy, $\exp((Nf^*_n(T)+Zf^*_p(T))/T)$, and binding energy, $\exp(-Q/T)$. Here, $f^*_{n(p)}(T)$ are the excited free energy per neutron (proton) of the source, and $Q$ is the $Q$-value associated with the emission. The degeneracy factors favor $^{11}$C. The values for $f^*$ are negative and hence the isotope with fewer nucleons is also favored by this factor. The magnitude of $f^*$, however, goes to zero like $T^2$ so the relative advantage for $^{11}$C arising from this factor deceases as the temperature goes down. The $Q$ values are greatly influenced by the respective binding energy factor, which strongly favors the $N=Z$ isotopes; this preference strengthens at reduced temperature. The net effect is that the preference for the $^{11}$C at the highest temperatures shifts to $^{12}$C as the temperature falls with the emission and expansion. We tested whether the binding energy is the dominant consideration by forcing the binding energies for $^{11}$C and $^{12}$C to be equal. In this test calculation, the relative emission rates for the two isotopes changed little with time.

In the EES model, $^{11}$C is preferentially emitted earlier than $^{12}$C, and the shapes of the energy spectra are consequently not the same. In Fig. 1, we show the calculated energy spectra for $^{11}$C and $^{12}$C (dashed lines) by taking integrating the spectra for emission from the time evolving source over time. The EES model correctly predicts the energy spectrum for $^{11}$C will be shifted to higher energies than that for $^{12}$C, and describes the higher energy tails of the spectra better than do the SMM calculations; nonetheless the slope of the spectra for both isotopes is still somewhat underpredicted. We note that it is necessary to multiply the EES model predictions for both $^{11}$C and $^{12}$C by a factor of 0.3 to match them to the data. We attribute this reduction to the fact that the emission of elements with $Z>10$ are not considered in these EES model calculations.

The total yield of $^{12}$C contains contributions from the neutron decay of excited $^{13}$C and the α decay of excited $^{16}$O. The yields of $^{11}$C are not affected significantly by sequential decays. Integrating over the energy spectra, we find an average kinetic energy of about 56.7 MeV for $^{11}$C and 45.2 MeV for $^{12}$C. The difference of about 11 MeV is in qualitative agreement with the data. The calculation predicts a larger fraction (≈23%) of the $^{11}$C fragments, a smaller fraction (≈7%) of the $^{12}$C fragments and an even smaller (<6%) of the heavier (A>12) carbon isotopes are emitted from the surface before the central density of the system expands below 0.4 $\rho_0$ where bulk disintegration occurs. By such low densities, the system has cooled to temperatures, T≤6 MeV, and it continues cooling to lower temperatures where the emission of the poorly bound neutron deficient $^{11}$C is relatively suppressed. This same scenario applies to the other elements, each of which shows similar patterns for relative emission. We show the EES calculations for the mean energies of all isotopes in the middle panel of Fig. 2. The EES model reproduces the basic trends of the data well.

In addition to these results, other considerations support the hypothesis of an early surface emission of fragments prior to the bulk disassembly of the expanded system. For lower incident energies, surface evaporation and fission are the basic decay modes of equilibrated compound nuclei, and for energies similar to the present study, transport theory predicts an abundant early emission of nucleons and clusters through the surface of expanding systems. The preequilibrium emission of fragments have been previously reported in proton induced reactions [23], and similar isotopic effects were observed. Here we have shown how to extract the significance of such effects for multifragmentation processes that have been widely interpreted as equilibrium bulk disintegration. Our results are qualitatively consistent with conclusion regarding the importance of surface fragment emission deduced previously from fragment-fragment correlation data [24]. They point one direction towards obtaining more comprehensive time dependent pictures of how fragmentation proceeds during the expansion of excited system.

In summary, we have shown that the measured energy spectra of the IMF isotopes produced in multifragmentation reveal the dynamics of the emission process. The more energetic neutron deficient isotopes are consistent with the picture that they are emitted earlier and more abundantly from the surface of the system while the source is expanding and cooling.

This work is supported by the National Science Foundation under Grant Nos. INT-0228058, PHY-01-10253, PHY-00-70818, PHY-00-70161 and by the DOE under grant numbers DE-FG02-87ER-40316 and by CNPq, Brazil under the contract No. 41.96.0886.00 of MCT/FINEP/CNPq (PRONEX).

**FIGURE CAPTIONS:**

Figure 1: The solid and open points represent the measured center of mass energy spectra for $^{12}$C and $^{11}$C fragments, respectively. The solid lines represent the corresponding ISMM calculations. The dashed lines represent the corresponding EES model calculations.

Figure 2: Left panel: experimental fragment mean kinetic energies. Middle panel: mean kinetic energies calculated with the EES model. Right panel: mean kinetic energies calculated with the ISSM model.

Figure 3: Left panel: Mean center of mass kinetic energies for $^{12}$C ($^{11}$C) calculated as a function of time with the EES model. Lower right panel: Emission rates for $^{12}$C ($^{11}$C) calculated as a function of time with the EES model. Upper right panel: Ratio of the emission rate for $^{11}$C divided by the emission rate for $^{12}$C calculated with the EES model

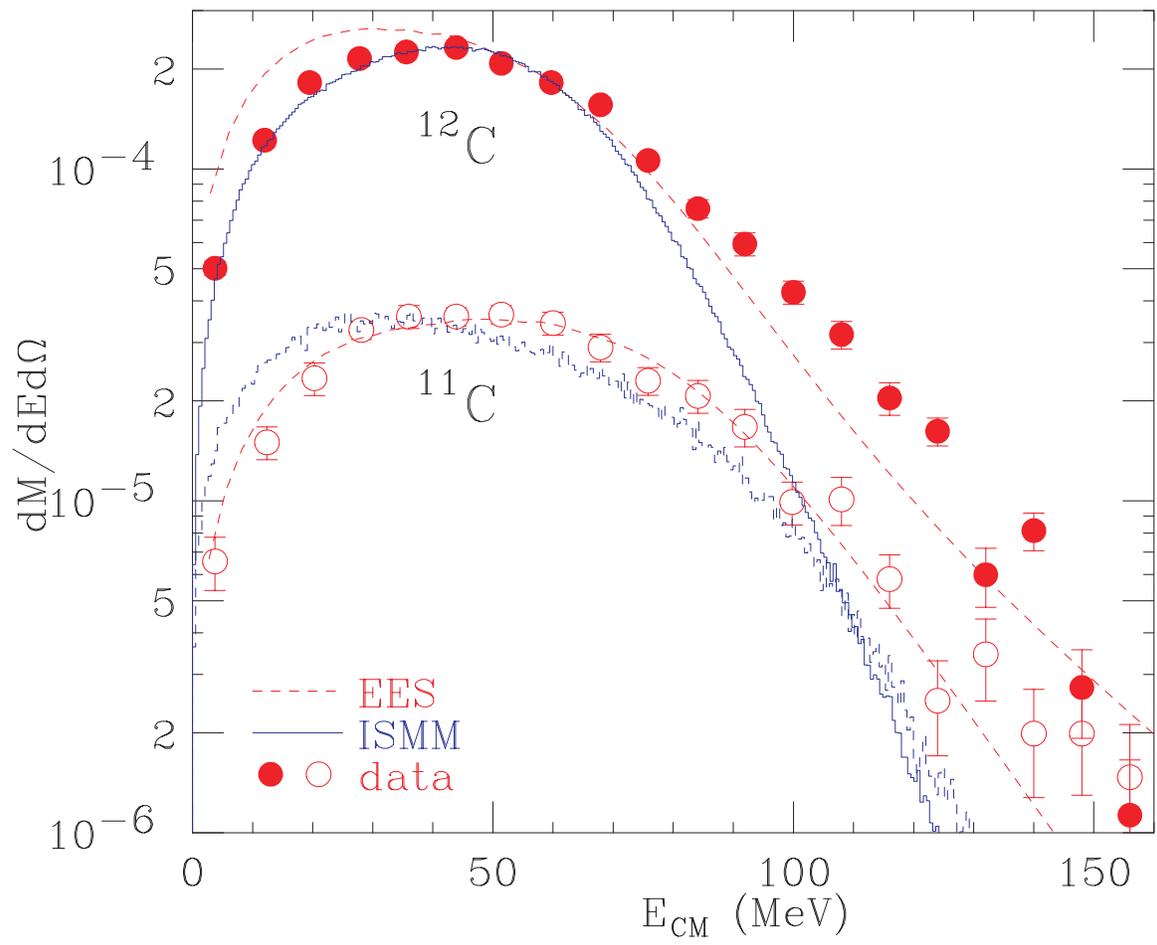

Fig. 1

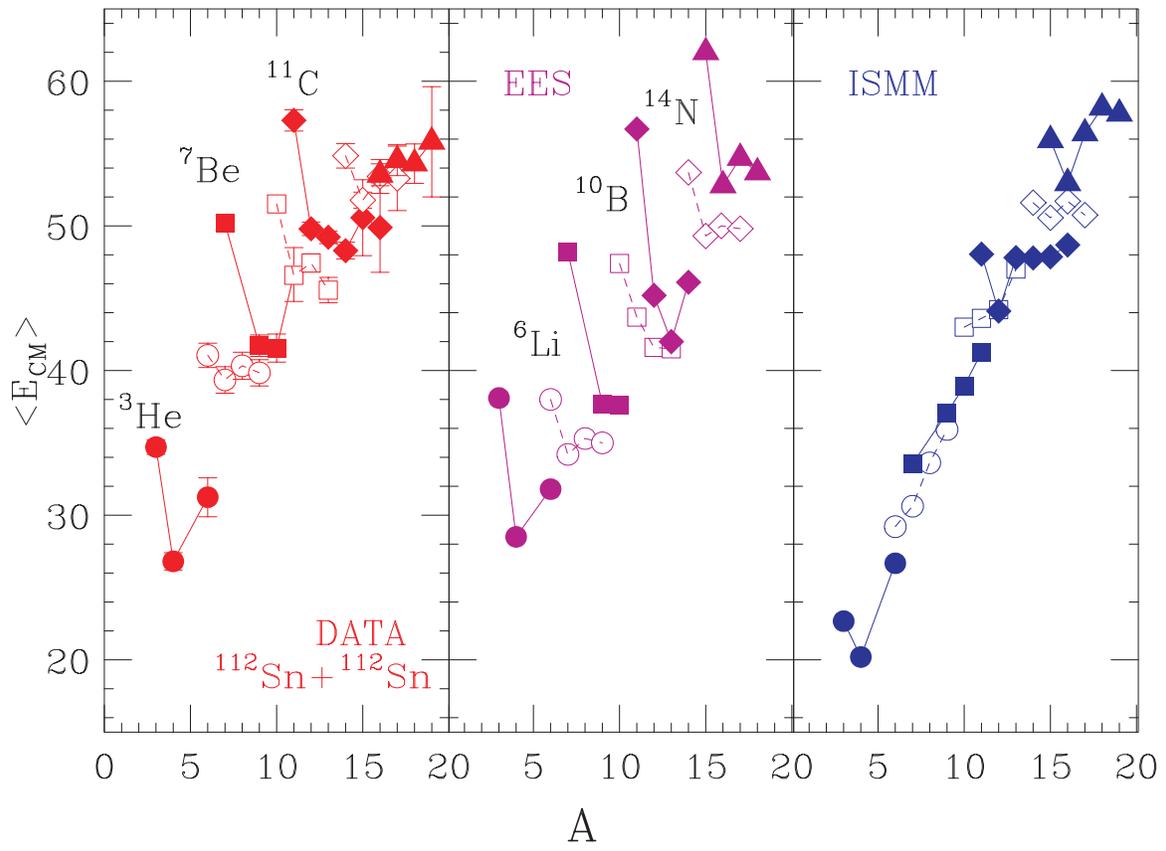

Fig. 2

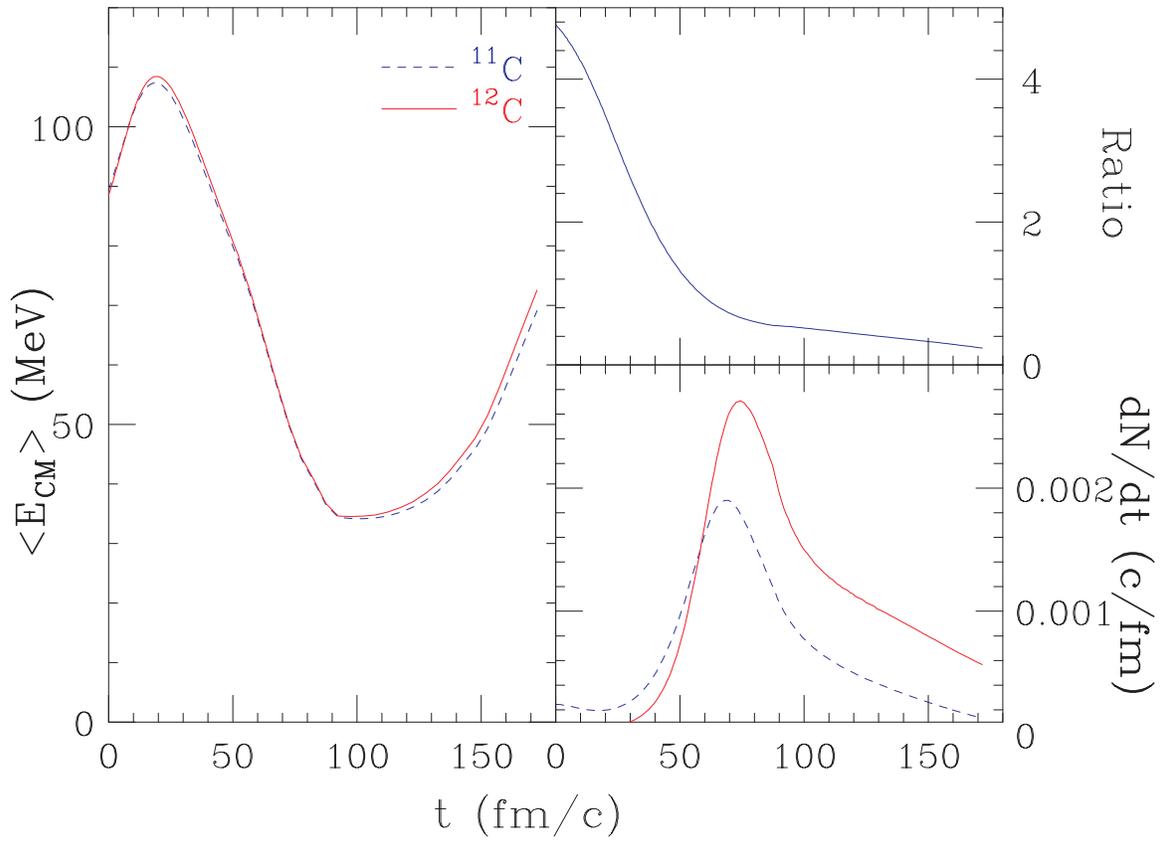

Fig. 3